\documentclass[preprint]{rsl}

\title{SAGECal performance with large sky models}
\author{Hanno Spreeuw, Sarod Yatawatta, Ben van Werkhoven and Faruk Diblen}

\begin{document}

\maketitle

%
%

\begin{abstract}
 As astronomical instruments become more sensitive, the requirements for the calibration software become more stringent; without accurate calibration solutions, thermal noise levels in images will not be reached and the scientific output of the instrument is degraded. Calibration requires bright sources with known properties, in particular with respect to their brightnesses as a function of frequency. However, for modern radio telescopes with a huge field of view, a single calibration source does not suffice; instead a sky model with tens of thousands of sources is needed. In this work, we investigate the compute load for such complicated sky models, with up to $50,000$ sources, for the SAGECal calibration package. We have chosen half of the sources in these models to be point sources and half of them extended, which we represent by Gaussian profiles.
 
\end{abstract}

\section{Introduction}

From its beginning, in the 1930s, radio astronomy progressed almost half a century with the use of simple calibration techniques, i.e., by calibrating on very bright unresolved sources with known flux density. These "standard candles" were observed just before and/or after the observation of the target source. The gain solutions from the individual dishes as calculated from the calibrator observation were applied to the target observation before images were made. This obviously cannot account for any gain phase changes during the target observation and it will be hard, if not impossible, to reach thermal noise levels in the images in this way. During the 1980s a new calibration technique, known as self-calibration or hybrid mapping found its way to aperture synthesis arrays \cite{Readhead1978}, although it had been applied to VLBI data earlier \cite{Pearson1984}. During a self-calibration loop a sky model is produced from the observation itself; this is used for the next iteration to find more accurate gain solutions until these solutions converge and the rms noise in the image no longer decreases. 
It should be obvious that more complete representations of the radio sky, as probed by the interferometer, will lead to more accurate calibration solutions. The most complete representations, however, will consist of tens of thousands of extended sources and point sources. These large numbers of sources will lead to substantial compute loads; first when these complex sky models are Fourier transformed to the (u, v, w) spacings of the observation - in order to find the Jones matrices that provide the optimal mapping from these "ground truth" visibilities to observed visibilities - and second to compute the residuals after the Jones matrices have been derived. For this purpose, we will investigate the runtimes of the SAGECal calibration package \cite{Yatawatta2012, SAGECal_Github} for sky models of five different sizes, to find a single set of calibration solutions - we have not run any selfcal loops, since the sky model is considered perfect. We also vary the number of stations in the observations that we calibrate. In this way, we derive the scaling of SAGECal performance for increasingly sensitive radio telescopes, from LOFAR~\cite{LOFAR_url} to SKA~\cite{SKA_url}.

\section{Method}

We used the same set of five artificial observations as we did in earlier work~\cite{2019arXivSpreeuw}. These are mock observations with 64, 128, 256, 384 and 512 stations, more or less ranging from the number of stations in the International LOFAR Telescope~\cite{LOFAR_Haarlem, ILT} to SKA1-low~\cite{SKA1Low}. We ran SAGECal on a dual Intel Xeon E5-2660v3 (40 logical cores), with both a NVIDIA Titan X and GeForce 1080 GPU. SAGECal was set to use 40 CPU threads and both of the GPUs for sufficient workloads, without any preferred order for deploying the GPUs. In this work we have set the number of clusters, i.e., the number of calibration directions, to 1 for all runs. By keeping this dimension fixed it becomes easier to focus on the effect of the complexity of the sky model on SAGEcal runtimes. We chose half of the sources in each sky model to have Gaussian profiles and the other half to be point sources. The peak brightnesses, sky positions and spectral indices are randomly distributed between some limits. Likewise for the major axes and orientation of the Gaussians, with the size of the minor axis always equal to half the size of the major axis. As in previous work~\cite{2019arXivSpreeuw}, we depict the times for finding the actual calibration solutions, by running SAGECal with both the "-a 0" and the "-a 1" settings. Running with the latter setting will Fourier transform the sky model to visibility space, while running with the former also includes the time to derive calibration solutions. Hence, the assessment of SAGECal performance in this work is built on the assumption that the Fourier transform of the sky model can be kept in RAM, which may not be possible if the observation has many frequency channels. 
As in our earlier work, we simultaneously investigate the effect of an increasing number of antenna stations on the time to compute calibration solutions. Although SAGECal can process a single observation on multiple nodes, we do not consider that here. It was, however,  discussed elsewhere~\cite{Data_multiplexing, Distributed_calibration} within the context of consensus optimization.

\section{Results}

\begin{figure}
	\centering
	\includegraphics[width=\columnwidth]{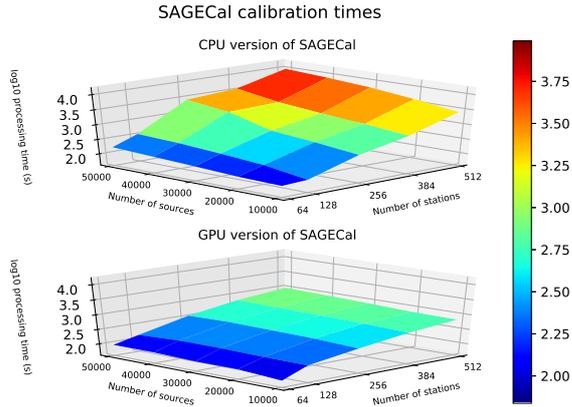}
	\caption{We measured the times SAGECal needs to calibrate five artificial data sets, having 64, 128, 256, 384 and 512 stations, for five different sky models, with varying numbers of sources, but all in a single cluster (calibration direction). We have used a computer equipped with two CPUs (2x Xeon E5-2660v3, 40 logical cores, with number of CPU threads = 40 in all runs), equipped with a NVIDIA GeForce GTX 1080 and Titan X GPU on ASTRON's DAS5, see~\cite{DAS5}.}
	\label{fig:3D-results}
\end{figure}

Figure~\ref{fig:3D-results} shows the results of our SAGECal runs for radio telescopes and sky models of five different sizes. It is clear from this figure that the GPU version of SAGECal scales better for increasing workloads. While the performance is similar for the lowest workload - 99s for the GPU version of SAGECal vs. 69s for its CPU version - SAGECal on a GPU is 10 times faster than on a CPU for the heaviest workload, as we expected.

\begin{figure}
	\centering
	\includegraphics[width=\columnwidth]{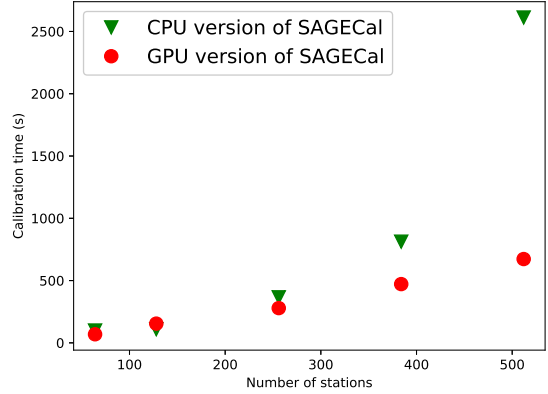}
	\caption{Slice from Figure~\ref{fig:3D-results} to show how the SAGECal runtimes are affected - on both CPU and GPU - by the size of the radio telescope - from LOFAR to SKA - when calibrating on our smallest sky model.}
	\label{fig:2D-results-10000-sources}
\end{figure}

\begin{figure}
	\centering
	\includegraphics[width=\columnwidth]{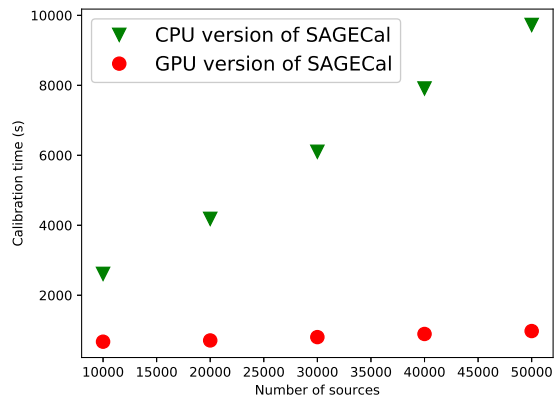}
	\caption{Slice from Figure~\ref{fig:3D-results} to show how the complexity of sky models affects the time SAGECal needs to derive calibration solutions, for its CPU and GPU version.}
	\label{fig:2D-results}
\end{figure}

\begin{figure}
	\centering
	\includegraphics[width=\columnwidth]{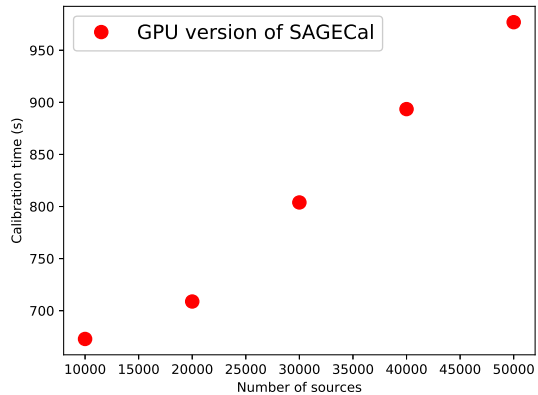}
	\caption{A zoom in on the GPU values from Figure~\ref{fig:2D-results} for quantative insight in the effect of larger sky models on SAGECal runtimes.}
	\label{fig:2D-results-GPU-only}
\end{figure}

Figure~\ref{fig:2D-results-10000-sources} shows SAGECal scalability - from LOFAR to SKA - for our smallest sky model: $1e4$ sources. While the runtimes of the CPU version of SAGECal are comparable to the GPU version for 64 and 128 stations, it shows scaling issues for observations from even larger telescopes. The GPU version, on the other hand, shows close to linear scaling over the entire range of observations.

Figure~\ref{fig:2D-results} is another slice of the 3D plot from Figure~\ref{fig:3D-results}, but in the orthogonal dimension; across all sky models for 512 stations. It shows linear scaling for both devices, but the slope is much shallower for the GPU than for the CPU, resulting in a factor ten difference in processing time for a sky model with $5e4$ sources.

Figure~\ref{fig:2D-results-GPU-only} shows only the GPU values from Figure~\ref{fig:2D-results} to get a more accurate understanding of this SAGECal runtimes on a typical SKA1-low observation using large sky models. It is striking that, when comparing a relatively small sky model with 5000 extended sources and 5000 point sources to a sky model five times larger, the time to compute calibration solutions only increases by 45\%. This is a promising result with regard to accurate calibration of large observations, such as the observations that are expected to be performed with SKA1-low.

\section{Conclusions}

In order to foresee possible compute bottlenecks in accurately calibrating data from increasingly large and sensitive radio telescopes we have investigated SAGECal runtimes using mock observations. As expected from earlier work, we notice insufficient linear scaling from a typical LOFAR observation to a SKA1-low observation even when calibrating on a relatively small sky model - consisting of $5e3$ point sources and $5e3$ extended sources - on a CPU with 40 logical cores. The GPU version does show linear scaling across all telescope sizes. With regard to SAGECal runtimes in the orthogonal dimension - from sky models with $1e4$ to $5e4$ sources, for an SKA1-low observation - we see linear scaling on both the CPU and the GPU - with a much shallower slope for the latter. This is reassuring considering the most accurate and therefore the most compute-intensive and data-intensive calibration ever that is coming up with the start of the early science observations of the SKA. 

\bibliographystyle{IEEEtran}
\bibliography{IEEEabrv, refs}

%
%
%
%
%
%
%

%
%
\noindent\small
Hanno Spreeuw is with Netherlands eScience Center, Science Park 140, Amsterdam, The Netherlands; e-mail: h.spreeuw@esciencecenter.nl.\\[1ex]
Sarod Yatawatta is with ASTRON, The Netherlands Institute for Radio Astronomy, Dwingeloo, The Netherlands; e-mail: yatawatta@astron.nl.\\[1ex]
Ben van Werkhoven is with Netherlands eScience Center, Science Park 140, Amsterdam, The Netherlands; e-mail: b.vanwerkhoven@esciencecenter.nl.\\[1ex]
Faruk Diblen is with Netherlands eScience Center, Science Park 140, Amsterdam, The Netherlands; e-mail: f.diblen@esciencecenter.nl.\\

\end{document}